\documentclass[aps,prc,showpacs,preprint,preprintnumbers,
showkeys,tighten,floatfix,nofootinbib]{revtex4}
\usepackage{amsmath,amssymb,amsfonts}
\usepackage{graphicx}
\usepackage{subfigure}
\usepackage{color}
\allowdisplaybreaks

\begin{document}

\title{ Conserved Density Fluctuation and Temporal 
Correlation Function in HTL Perturbation Theory}

\author{Najmul \surname{Haque}$^a$}
\author{Munshi G. \surname{Mustafa}$^{a}$}
\author{Markus H. \surname{Thoma}$^b$}
\affiliation{$^a$Theory Division, Saha Institute of Nuclear Physics,
1/AF Bidhannagar, Kolkata 700 064, India }
\affiliation{$^b$Max-Planck-Institut f\"ur extraterrestrische Physik,
Giessenbachstrasse, 85748 Garching, Germany}

\begin{abstract}
Considering recently developed Hard Thermal Loop perturbation 
theory that takes into account the effect of the variation of the 
external field through the fluctuations of a conserved quantity 
we calculate the temporal component of the Euclidian correlation function 
in the vector channel. The results are found to be in good agreement with the 
very recent results obtained within the quenched approximation of QCD and 
small values of the quark mass ($\sim 0.1T$)  on improved lattices 
of size $128^3\times N_\tau$ at ($N_\tau=40, \ T=1.2T_C$),
($N_\tau=48, \ T=1.45T_C$), and ($N_\tau=16 , \ T=2.98T_C$), where $N_\tau$
is the temporal extent of the lattice.
This suggests that the results from lattice QCD and Hard Thermal Loop  
perturbation theory are in close proximity for a quantity associated 
with the conserved density fluctuation.
\end{abstract}

\pacs{12.38.Cy, 12.38.Mh, 11.10.Wx}
\keywords{Quark-Gluon Plasma, Hard Thermal Loop Approximation,
Conserved Charges, Correlation Function}
%\preprint{SINP/TNP/2007-31}
\maketitle

\section{Introduction}
Dynamical properties of many particle system can generally be studied by
employing an external perturbation, which disturbs the system only slightly
from its equilibrium state, and thus measuring the spontaneous 
response/fluctuations of the system to this external perturbation.
In general, the fluctuations are related to the correlation function through
the symmetry of the system, which provides important inputs for quantitative 
calculations of complicated many-body system. Also, many of the properties 
of the deconfined strongly interacting matter are reflected in the structure 
of the correlation and the spectral functions~\cite{hasimoto} of the 
vector current.

The static thermal dilepton rate describing the production of lepton pairs is
related to the spectral function in the vector current~\cite{munshi1,karsch}. 
Within the Hard thermal loop perturbation theory (HTLpt) the vector spectral 
function has been obtained~\cite{munshi1,beradau,moore}, which is found 
to diverge due to its spatial part at the low 
energy regime. This is due to the fact that the HTL quark-photon vertex 
is inversely proportional to the photon energy and it sharply rises at 
zero photon energy. On the other hand, the fluctuations of conserved 
quantities, such as baryon number and electric charge, are considered 
to be a signal~\cite{expt,stephanov} for quark-gluon plasma (QGP) 
formation in heavy-ion experiments. These conserved density fluctuations 
are closely related to the 
temporal correlation function in the vector channel through derivatives 
of a thermodynamic quantity associated with the symmetry, known as 
the thermodynamic sum rule~\cite{calen}. It is expected that the 
temporal part of the spectral function associated with the symmetry 
should be a finite quantity and would not encounter any such infrared 
divergence unlike the spatial part at low energy.  A very recent lattice 
calculation~\cite{karsch2} in quenched approximation has obtained the 
temporal part of the Euclidian
correlation function associated with the response of the conserved 
density fluctuations, which is found to be a finite quantity. In view 
of this we would like to compute the
temporal correlation function in the vector channel from the quark number 
susceptibility associated with the quark number density fluctuations 
within the HTLpt~\cite{munshi2} and to compare it with recent lattice 
data~\cite{karsch2} in quenched approximation. 

The temporal correlator follows from the QCD polarization diagram. To lowest 
order perturbation theory it is given by the one-loop diagram containing bare 
quark propagators. The HTL resummation technique provides a consistent 
perturbative expansion for gauge theories at finite temperature by using 
HTL resummed propagators and vertices \cite{braaten}. As in usual perturbation 
theory it is strictly applicable only in the weak coupling limit but takes 
into account all dynamical effects. Going beyond the lowest order perturbation 
theory in the case of the temporal correlator, we use HTL resummed quark 
propagators and quark-gluon vertices in the polarization diagram 
(see \cite{munshi2,munshi}). The HTL resummed quark propagators correspond to 
static external quarks (valence quarks). Within this approximation no internal 
quark-loops appear. In this sense our approximation can be compared to results 
from quenched lattice QCD. The inclusion of dynamical quark-loops requires to 
consider higher-order diagrams within the HTL resummed perturbation theory in 
which HTL resummed gluon propagators will show up. However, we do not expect 
a significant change in the result as higher contributions give less than 5\% 
corrections in the case of thermodynamic quantities such as pressure 
\cite{andersen1}. Of course, close to the transition temperature higher order 
effects may become important~\cite{andersen2}.      
  
The paper is organised as follows. In sec.II we briefly discuss some
generalities on correlation functions,  fluctuation and its response
(susceptibility) associated with conserved charges.
In sec.III we obtain the relation between the response of the density
fluctuation of the conserved charge and the corresponding temporal part of 
the Euclidian correlation function in the vector current. Next we compute them in 
HTLpt~\cite{munshi2} and compare with lattice data.  Finally, we conclude 
in sec.VI.

\section{Generalities}
In this section we summarize some of the basic relations and also 
describe in details their important features relevant as well as 
required for our purpose.

\subsection{Correlation Functions}

The two-point correlation function~\cite{hasimoto,munshi1,karsch} of 
the vector current, 
$J_\mu={\bar {\psi}}(\tau,{\vec {\mathbf x}})\Gamma_\mu \psi(\tau,{\vec
{\mathbf x}})$ with three point function $\Gamma_\mu$, 
%which can be calculated numerically in the framework of lattice gauge theory 
is defined at fixed momentum ${\vec {\mathbf p}}$ as 
\begin{eqnarray}
G_{\mu\nu}(\tau,{\vec{\mathbf p}})=\int d^3x \ \langle J_\mu(\tau, 
{\vec {\mathbf x}}) J_\nu^\dagger(0, {\vec {\mathbf 0}})\rangle \ 
e^{i{\vec{\mathbf p}}\cdot {\vec{\mathbf x}}} \ , \label{cor_p}
\end{eqnarray}
where the Euclidian time $\tau$ is restricted to the interval
$[0,\beta=1/T]$. The thermal two-point vector correlation function 
in coordinate space, $ G_{\mu\nu}(\tau, {\vec {\mathbf x}})$, can be written as
\begin{equation}
G_{\mu\nu}(\tau, {\vec {\mathbf x}})=\langle J_\mu(\tau, {\vec {\mathbf x}})
J_\nu^\dagger(0, {\vec {\mathbf 0}})\rangle =T\sum_{n=-\infty}^{\infty} \int
\frac{d^3p}{(2\pi)^3} \ e^{-i(w_n\tau+{\vec{\mathbf p}}
\cdot {\vec{\mathbf x}})}\ G_{\mu\nu}(w_n,{\vec{\mathbf p}}) \ , 
\label{vec_cor}
\end{equation}
where the Fourier transformed correlation function 
$G_{\mu\nu}(w_n,{\vec{\mathbf p}})$ is given at the discrete
Matsubara modes, $w_n=2\pi n T$.  The imaginary part of the momentum space
correlator gives the spectral function $\sigma(\omega,{\vec{\mathbf p}})$ as
\begin{eqnarray}
G_H(w_n,{\vec{\mathbf p}}) &=& - \int_{-\infty}^{\infty} d\omega
\frac{\sigma_H(\omega,{\vec{\mathbf p}})}{iw_n-\omega+i\epsilon} \nonumber \\ 
\ \Rightarrow \
\sigma_H(\omega,{\vec{\mathbf p}})&=& \frac{1}{\pi} {\mbox{Im}}
\ G_H(\omega,{\vec{\mathbf p}}) \ , \label{spec_mom}
\end{eqnarray}
where $H=(00,ii,V)$ denotes (temporal, spatial, vector). We have 
also introduced the vector spectral function 
as $\sigma_V=\sigma_{00}+ \sigma_{ii}$, where $\sigma_{ii}$ is the sum 
over the three space-space components and $\sigma_{00}$ is the time-time 
component of $\sigma_{\mu\nu}$.

Using (\ref{vec_cor}) and (\ref{spec_mom}) in (\ref{cor_p}) 
the spectral representation of
the thermal correlation functions at fixed momentum 
can be obtained~\cite{munshi1} as
\begin{equation}
G_H(\tau, {\vec {\mathbf p}})=\int_0^\infty\ d\omega \
\sigma_H (\omega, {\vec {\mathbf p}}) \
\frac{\cosh[\omega(\tau-\beta/2)]}{\sinh[\omega\beta/2]} \ . \label{corr_mom}
\end{equation}
We note that the Euclidian correlation function is
usually restricted to vanishing three momentum, $\vec{\mathbf p}=0$, 
in the analysis of lattice gauge theory and one can write 
$G_H(\tau T)=G_H(\tau,{\vec{\mathbf 0}})$.  

A finite temperature 
lattice gauge theory calculation is performed on
lattices with finite temporal extent $N_\tau$, which provides information
on the Euclidian correlation function, $G_H(\tau T)$, only for a discrete 
and finite set of Euclidian times 
$\tau =k/(N_\tau T), \ \ k=1,\cdots \ N_\tau$. The vector correlation 
function, $G_V(\tau T)$, had been computed~\cite{karsch} within the 
quenched approximation of QCD\footnote{In comparison to thermodynamic 
quantities some quantities like mesonic correlation and spectral functions 
due to their structures are still too exhaustive and expensive to 
calculate in full QCD with improved lattice action. Hence the quenched 
approximation is still very useful to understand the various features of 
correlation and spectral functions.}
using
non-perturbative improved clover fermions~\cite{clov} through a probabilistic
application based on the maximum entropy method (MEM)~\cite{mem} for
temporal extent $N_\tau= 16$ and spatial extent $N_\sigma = 64$. Then by 
inverting the integral in (\ref{corr_mom}), the spectral function\footnote{We
note that lattice technique cannot directly be used to obtain the spectral 
function due to its difficulty to perform an analytic continuation of 
(\ref{spec_mom}) from imaginary time to real time.} was 
reconstructed~\cite{karsch,aarts} in lattice QCD. The vector spectral functions 
above the deconfinement temperature ({\em viz.}, $T=1.5T_c \ {\mbox{and}} 
\ 3T_c$) show an oscillatory behaviour compared to the free one. In the high energy
regime, $\omega/T\ge 4$ the vector spectral function, $\sigma_V(\omega,
{\vec{\mathbf 0}})$ agreed with
that of the HTLpt~\cite{munshi1,carsten}. On the other hand, the 
lattice spectral 
functions and dilepton rates~\cite{karsch} were found to fall off very 
fast and became vanishingly small for $\omega/T\le 4$ due to the sharp 
cut-off used in the reconstruction.

In a very recent lattice analysis~\cite{karsch2} the low energy 
behaviour ({\it viz.}, downward slope) of the spatial and vector spectral 
functions has been improved substantially on lattices up to size
$128^3\times 48$ by changing the slope upward through a fit to the free 
plus Breit-Wigner (BW) spectral functions at lower energy limit. This upward 
slope in the vector spectral functions in lattice gauge theory~\cite{karsch2} 
resembles up to some extent that of the HTLpt spectral 
function~\cite{munshi1,beradau} 
and dilepton rate~\cite{moore,carsten} at low energy regime despite the 
infrared problem of HTLpt at vanishing energy. Nonetheless, the existence 
of the van Hove peaks in the vector spectral function 
in HTLpt~\cite{munshi1,yaun,markus, wong} has not
been realized yet in the improved lattice analysis~\cite{karsch2}, probably 
due to the ansatz that the lattice spectral function is proportional to 
that of the free plus BW one in the low energy regime. The existence of
van Hove peaks cannot yet be ruled out, which are general features of 
massless fermions~\cite{yaun,markus,wong} in a relativistic plasma in 
the low energy regime. Also, the high energy behaviour of 
the spectral function agrees well with those of free and HTLpt results, 
respectively, and are  in conformity with its earlier 
analysis on the lattice~\cite{karsch}. On the other hand, the temporal component of 
spectral and correlation functions associated with the symmetry of the system
are finite and do not encounter any 
infrared problem unlike their spatial part. Now, it would be interesting 
to analyse the temporal correlation and spectral functions, associated 
with the symmetry, within HTLpt and compare them with those of lattice 
gauge theory~\cite{karsch2}.

\subsection{ Density Fluctuation and its Response}

Let \( {\cal O}_{\alpha } \) be a Heisenberg operator where $\alpha$ 
may be associated with a degree of freedom in the system.
In a static and uniform
external field \( {\cal F}_{\alpha } \), the (induced)
expectation value of the operator \( {\cal O}_\alpha \left( 0,
\overrightarrow{\mathbf x}
\right) \) is written~\cite{calen,kunihiro} as
\begin{equation}
\phi _{\alpha }\equiv \left\langle {\cal O} _{\alpha }\left
( 0,\overrightarrow{\mathbf x}\right) \right\rangle _{\cal F} = \frac{{\rm Tr}\left
[ {\cal O} _{\alpha }\left( 0,\overrightarrow{\mathbf x}\right) e^{-\beta \left
( {\cal H}+{\cal H}_{ex}\right) }\right] }{{\rm Tr}\left[ e^{-\beta
\left( {\cal H}+{\cal H}_{ex}\right) }
\right] }=\frac{1}{V}\int d^{3}x\, \left\langle {\cal O} _{\alpha }
\left( 0,\overrightarrow{\mathbf x}\right) \right\rangle \: , \label{eq1}
\end{equation}
where translational invariance is assumed, $V$ is the volume of the
system and
\({\cal H}_{ex} \) is given by
\begin{equation}
{\cal H}_{ex}=-\sum _{\alpha }\int d^{3}x\, {\cal O} _{\alpha }\left( 0,
\overrightarrow{\mathbf x}\right) {\cal F}_{\alpha }\: .\label{eq2}
\end{equation}

The (static) susceptibility \( \chi _{\alpha \sigma } \) is defined as
the rate with which the expectation value changes
in response to that external field,
\begin{eqnarray}
\chi _{\alpha \sigma }(T) & = & \left. \frac{\partial \phi _{\alpha }}
{\partial {\cal F}_{\sigma }}\right| _{{\cal F}=0}
  = \beta \int d^{3}x\, \left\langle {\cal O} _{\alpha }\left
( 0,\overrightarrow{\mathbf x}\right) {\cal O} _{\sigma }
( 0,\overrightarrow{\mathbf 0})
 \right\rangle \: , \label{eq3}
\end{eqnarray}
where
$\langle {\cal O}_\alpha (0,{\vec {\mathbf x}})
{\cal O}_\sigma(0,{\vec {\mathbf 0}})\rangle $
is the two point correlation function with operators evaluated
at equal times. There is no broken symmetry as
\begin{equation}
\left.\left\langle {\cal O} _{\alpha }
\left ( 0,\overrightarrow{\mathbf x}\right ) \right\rangle
\right|_{{\mathcal F}\rightarrow 0}
 =\left. \left\langle {\cal O} _{\sigma } 
( 0,\overrightarrow{\mathbf 0}) \right\rangle 
\right |_{{\mathcal F}\rightarrow 0}=0  \ . \label{eq3i}
\end{equation}

\section{{Quark Number Susceptibility (QNS) and Temporal Euclidian 
Correlation Function:}}

The QNS is a measure of the response of the quark number density with 
infinitesimal change in the quark chemical
potential, $\mu+\delta\mu $. Under such a situation the external field, 
${\cal F}_\alpha$, in ({\ref{eq2}}) can be identified as the  
quark chemical potential and the operator ${\cal O}_\alpha$ as
the temporal component ($J_0$) of the vector current, 
$J_\sigma(t,{\vec {\mathbf x}})= \overline{\psi} \Gamma_{ \sigma}\psi$, 
where $\Gamma^\sigma$ is in general a three point function.
Then the QNS for a given quark flavour follows from
(\ref{eq3}) as
\begin{eqnarray}
\chi_q(T) &=& \left.\frac{\partial \rho}{\partial \mu}\right |_{\mu=0}
= \left.\frac{\partial^2 {\cal P}}{\partial \mu^2}\right |_{\mu=0}
= \int \ d^4x \ \left \langle J_0(0,{\vec {\mathbf x}})J_0(0,{\vec {\mathbf 0}})
\right \rangle \ =- {\lim_{\vec{\mathbf p}\rightarrow 0}} 
{\mbox {Re}}\ G_{00}^R(0,{\vec{\mathbf p}}),
\label{eq4}
\end{eqnarray}
where $G_{00}^R$ is the retarded correlation function. To obtain 
(\ref{eq4}) in concise form, we have used the fluctuation-dissipation 
theorem given as
\begin{equation}
G_{00}(\omega,{\vec{\mathbf p}})=-\frac{2}{1-e^{-\omega/T}} {\mbox{Im}}
G_{00}^R(\omega,{\vec{\mathbf p}}), \label{eq4a}
\end{equation}
and the Kramers-Kronig dispersion relation
\begin{equation}
{\mbox{Re}}G_{00}^R(\omega,{\vec{\mathbf p}})=\int_{-\infty}^{\infty} 
\frac{d\omega^\prime}{2\pi}
 \frac{{\mbox{Im}}G_{00}^R(\omega,{\vec{\mathbf p}})}{\omega^\prime-\omega},
\label{eq4b}
\end{equation}
where $\lim_{\vec{\mathbf p}\rightarrow 0}{\mbox{Im}} G_{00}^R
(\omega,{\vec{\mathbf p}})$ is proportional to
$\delta(\omega)$ due to the quark number conservation~\cite{calen,kunihiro}. 
Also the number density for a given quark flavour can be written  as
\begin{equation}
\rho=\frac{1}{V}
\frac{{\rm{Tr}}\left [  {\cal N} e^{-\beta \left({\cal H}-\mu {\cal N}
\right )}\right ]}
{{\rm{Tr}}\left [e^{-\beta \left({\cal H}-\mu {\cal N}\right )}\right ]} \frac{\langle {\cal N}\rangle}{V} = \frac{\partial {\cal P}}
{\partial \mu} \ , \label{eq5}
\end{equation}
with the quark number operator, ${\cal N}=\int J_0(t, {\vec x}) \ d^3x
=\int {\bar \psi}(x)\Gamma_0\psi(x)d^3x$, and
${\cal P}=\frac{T}{V} \ln {\mathcal Z}$ is the pressure and 
${\mathcal Z}$ is the partition
function of a quark-antiquark gas. The quark number density vanishes
if $\mu\rightarrow 0$, i.e., there is no broken CP symmetry. 
Now, (\ref{eq3}) or (\ref{eq4}) indicates that the thermodynamic derivatives
with respect to the external source
are related to the temporal component of the static correlation function
associated with the number conservation of the system. This relation 
in (\ref{eq4}) is known as {\em the thermodynamic sum rule}~\cite{calen}.

Owing to the quark number conservation the temporal spectral function 
$\sigma_{00}(\omega,{\vec{\mathbf 0}})$ in (\ref{spec_mom}) becomes
\begin{equation}
\sigma_{00}(\omega,{\vec{\mathbf 0}})=\frac{1}{\pi} {\mbox{Im}} G_{00}^R
(\omega,{\vec{\mathbf 0}})= - \omega \delta(\omega) \chi_q(T) \, . \label{eq6}
\end{equation} 

Using (\ref{eq6}) in (\ref{corr_mom}), it is straight forward to obtain
the temporal correlation function as
\begin{equation}
G_{00}(\tau T)=-T\chi_q(T), \label{eq7}
\end{equation}
which is proportional to the QNS $\chi_q$ and $T$, but independent of $\tau$.
The QNS has been calculated within the framework of lattice 
gauge theory
\cite{lat1,lat3,peter,bazavov,milc,lat2,qmass,peter1}, perturbative
QCD~\cite{vuorinen}, Nambu-Jona-Lasinio (NJL) Model~\cite{kunihiro,fuji},
Polyakov-Nambu-Jona-Lasinio (PNJL) Model~\cite{ghosh},  Ads/CFT
correspondence and Holographic QCD~\cite{kim}, Renormalisation Group 
approach~\cite{schaefer},
two loop approximately self-consistent
$\Phi$-derivable  HTL resummation~\cite{blaizot}
and HTLpt~\cite{munshi2,munshi,jiang}. We note that  
in a resummed perturbation theory~\cite{pisarski,braaten}, the higher 
order loops contribute to the lower order due to the fact that the 
loop expansion and the coupling expansion are not symmetric. So, unlike 
conventional perturbation theory~\cite{kapusta} one needs to take a proper 
measure in order to calculate a quantity in a given order 
of $\alpha_s$ correctly using HTLpt.  We further note that various HTL
approaches~\cite{blaizot,munshi,jiang,andersen,blaizot1} have been
used for calculating LO thermodynamic quantities and 
QNS in the literatures, which led to different results within the same
approximation. Recently, we have developed a resummed HTLpt~\cite{munshi2} by 
employing a variation of an external probe that disturbs the system only 
slightly from its equilibrium positions. In this way the effect of higher 
order variations of the external source is taken into account, which are
essential to get the LO quantities correct. Once this is done the LO 
quantities in HTLpt~\cite{munshi2} agree with the HTL 
approaches of Ref.~\cite{blaizot,blaizot1}. 
Here we would like to exploit the LO QNS in 
HTLpt obtained in Ref.~\cite{munshi2} for our purpose. 

Because of the structure of the HTL propagator~\cite{pisarski,braaten} the  
LO QNS in HTLpt  contains quasiparticle (QP) contributions  due to the
poles of the HTL quark propagator and Landau damping (LD) contributions
due to the space like part of the HTL quark propagator. The QNS can be 
decomposed as
\begin{equation}
\chi_q^{HTL}(T)= \chi_{q}^{QP}+\chi_{q}^{LD}. \label{s5}
\end{equation}

The LO  QNS in HTLpt due to QP is obtained~\cite{munshi2}
as
\begin{eqnarray}
\chi_q^{QP}(T)
&=& 4N_cN_f\beta\int \frac{d^3k} {(2\pi)^3}\left [
n(\omega_+)\left(1-n(\omega_+)\right)
\ + \ n(\omega_-)\left(1-n(\omega_-)\right) \right. \nonumber \\
&&\left. \ \ \ \ \ \ \ \ \ \ \ \ \ \ \ \ \ \ \ \ \ \ \ \ \
-\ n(k)\left(1-n(k)\right)
\right ] \ , \label{H8}
\end{eqnarray}
and the LD part is obtained~\cite{munshi2} as
\begin{eqnarray}
\chi_q^{LD}(T)
=2N_cN_f\beta\int\frac{d^3k}{(2\pi)^3}
\int\limits_{-k}^k d\omega\left(\frac{2m_q^2}{\omega^2-k^2}\right)\
%\nonumber\\
%&&\ \ \ \times \ \
\beta_+(\omega,k)\ n(\omega)\left(1-n(\omega)\right) \ , \label{L4}
\end{eqnarray}
where $n(y)$ is the Fermi-Dirac distribution, $\omega_+$ corresponds to
the energy of a quasiparticle having chirality to helicity ratio $+1$, 
$\omega_-$ is the energy of a mode called plasmino having chirality to
helicity ratio $-1$, and $\beta_\pm$ are the cut spectral functions
of the HTL quark propagator. The QP part results in (\ref{H8}) 
is identical to that of the $2$-loop approximately self-consistent 
$\Phi$-derivable HTL resummation approach of Blaizot {\em et al}
~\cite{blaizot}. The LD part (\ref{L4}) cannot be compared directly to 
the LD part of Ref.~\cite{blaizot} as no closed expression is given 
there. However, numerical results of the both QNS agree very well.
We also note that Jiang {\em et al}~\cite{jiang} used 
HTLpt but did not take into account properly the effect of the variation 
of the external field to the density fluctuation, which resulted in 
an overcounting in the  LO QNS. Moreover, in their approach an ad hoc
separation scale is required to distinguish between soft and hard momenta and 
the thermodynamic sum rule is violated. 
In the HTLpt approach in Ref.~\cite{munshi}
the HTL N-point functions were used uniformly for all momenta scale,
{\em i.e.}, both soft and hard momenta, which resulted in an overcounting
within the LO contribution~\cite{blaizot}. The reason is that the HTL 
action is accurate only for
soft momenta and for hard ones only in the vicinity of light cone.

\begin{figure}[!tbh]
\begin{center}
\includegraphics[height=0.55\textwidth, width=0.65\textwidth]{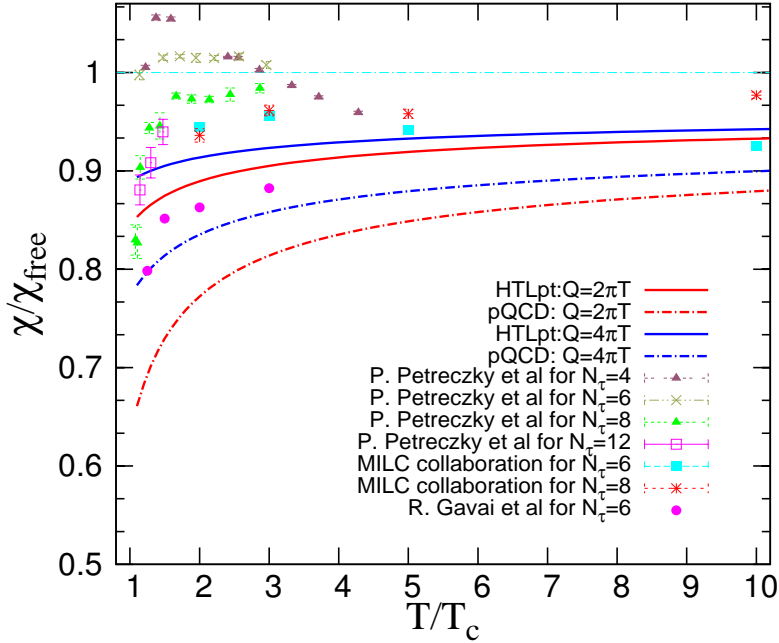}
%\vspace*{-0.3in}
\caption{(Color online) The 2-flavour scaled QNS with that of free
one as a function of $T/T_C$.
The solid lines are for LO in HTLpt whereas the dashed lines are 
for LO (proportional to $g^2$) in pQCD~\cite{blaizot,toimela,kapusta}. 
The different choices of the renormalisation
scale are $Q =2\pi T \ {\mbox{(red)}}, \ \
{\mbox{and}} \ \ 4\pi T \ {\mbox{(blue)}}$. The symbols represent the
various lattice data~\cite{lat3,peter,bazavov, milc,lat2}. 
The violet triangles (with $T_c=204\pm2$ MeV), brown crosses (with $T_c196\pm 3$ MeV), green triangles (with $T_c=191\pm 2$ MeV) and
purple squares (with $T_c=185 \pm 5$ MeV) represent
$p4$ lattice QCD data~\cite{peter,bazavov}.
% for (2+1)-flavour and the
%light quark mass
%$m=0.1m_s$, where $m_s$ is the strange quark mass near its physical value.
%We also note
%that the $T_c$ values for $p4$ action are unpublished and are based on
%estimates of the peak position of the chiral susceptibility from
%HotQCD collaboration~\cite{ranjan} for $N_{\tau}=8$, while for
%$N_{\tau}=12$ it is estimated based on the observed
%cutoff dependence of $T_c$ on $N_{\tau}=6$ and $N_{\tau}=8$ lattices.
The squares (cyan) and stars (saffron) are from asqtad lattice QCD
data~\cite{milc}.
% for (2+1)-flavour, $m= 0.2m_s$ and
%$T_c=186\pm 4$ MeV. 
The solid circles (purple) represent quenched QCD
data~\cite{lat2} for $T_c=0.49\Lambda_{\overline{MS}}$. 
The quark mass ranges between (0.1 to 0.2)$m_s$, 
where $m_s$ is the strange quark mass near its physical value.
% with 2-flavour, $m=0.1T_C$ and
%$T_c=0.49\Lambda_{\overline{MS}}$. 
Note that further lowering the quark
mass to its physical value seems to have a small effect~\cite{qmass} for
$T> 200$ MeV.  The details of these lattice results are also summarised 
in Ref.~\cite{peter1}. }
\label{qns_I}
\end{center}
\end{figure}

Now we display in Fig.~\ref{qns_I} the 2-flavour\footnote{We note
that the QNS has a very weak flavour dependence that enters through the
temperature dependence of the strong coupling as 
$\alpha_s(T)=\frac{12\pi}{(33-2N_f)\ln(Q^2/\Lambda_0^2)}$
where $Q$ is the momentum scale and $T_C=0.49\Lambda_0$.  }
scaled QNS in LO with that of free gas as a function of temperature that
shows significant improvement over pQCD results of order $g^2$
~\cite{blaizot,toimela,kapusta}. Moreover,
it also shows the same trend as the available lattice
results~\cite{lat2,lat3,peter,bazavov,milc,qmass,peter1}, though there
is a large variation among the various lattice results within the improved
lattice (asqtad and p4) actions~\cite{peter,milc} due to the higher order
discretisation of the relevant operator associated with the thermodynamic 
derivatives. A detailed analysis 
on uncertainties of the ingredients in the lattice QCD calculations is
presented in Refs.~\cite{bazavov,qmass}. This calls for further investigation 
both on the analytic side by improving the HTL resummation schemes and on 
the lattice side by refining the various lattice ingredients.

%\vspace{0.2in}
\begin{figure}[!tbh]
%\subfigure{
%{\includegraphics[scale=0.42]{M_chi.ps}}
{\includegraphics[height=0.48\textwidth, width=0.55\textwidth,
angle=270]{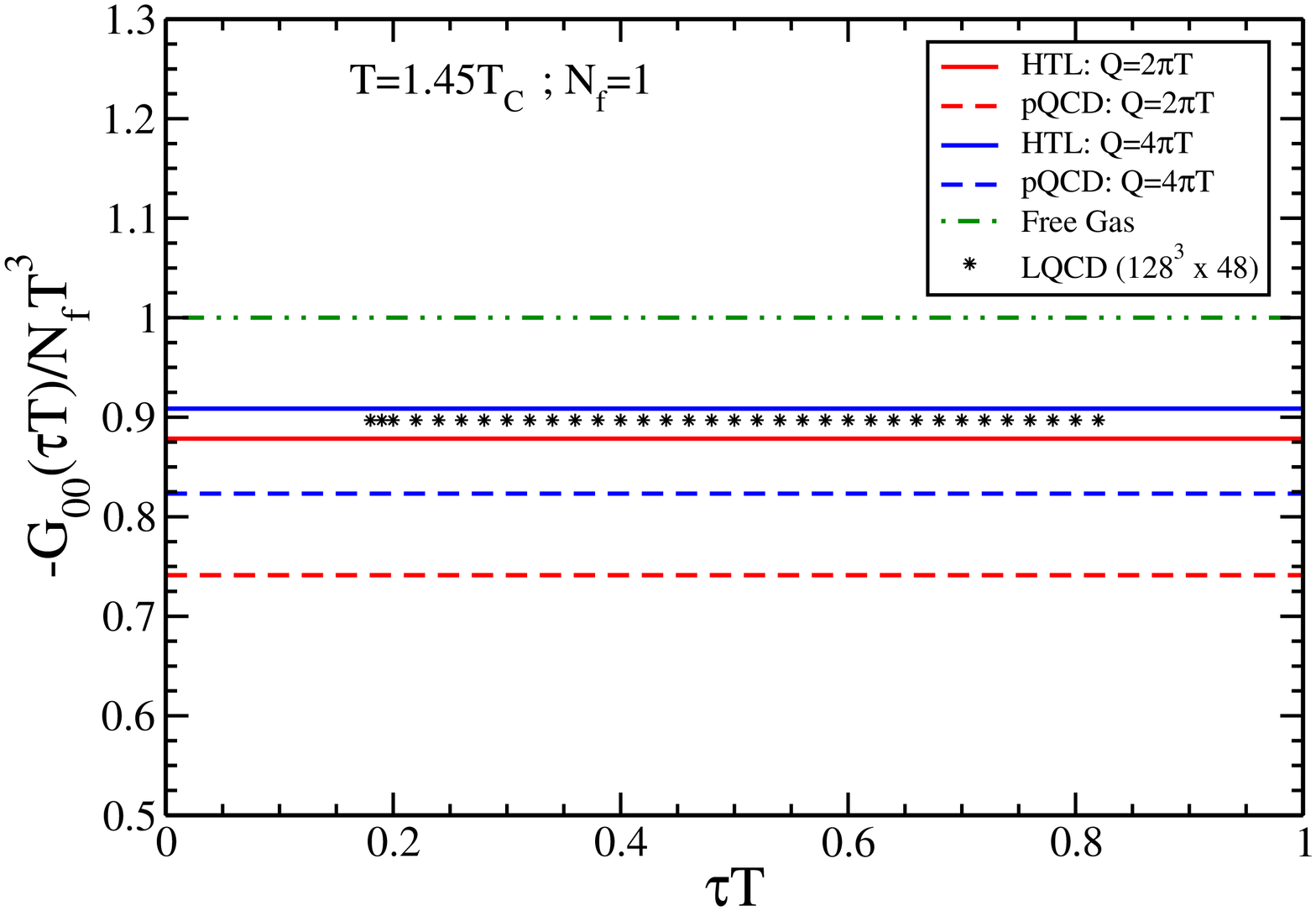}}
%}
%\vspace*{0.3in}
%\subfigure{
%{\includegraphics[scale=0.42]{M_R.ps}}
{\includegraphics[height=0.48\textwidth, width=0.55\textwidth,
angle=270]{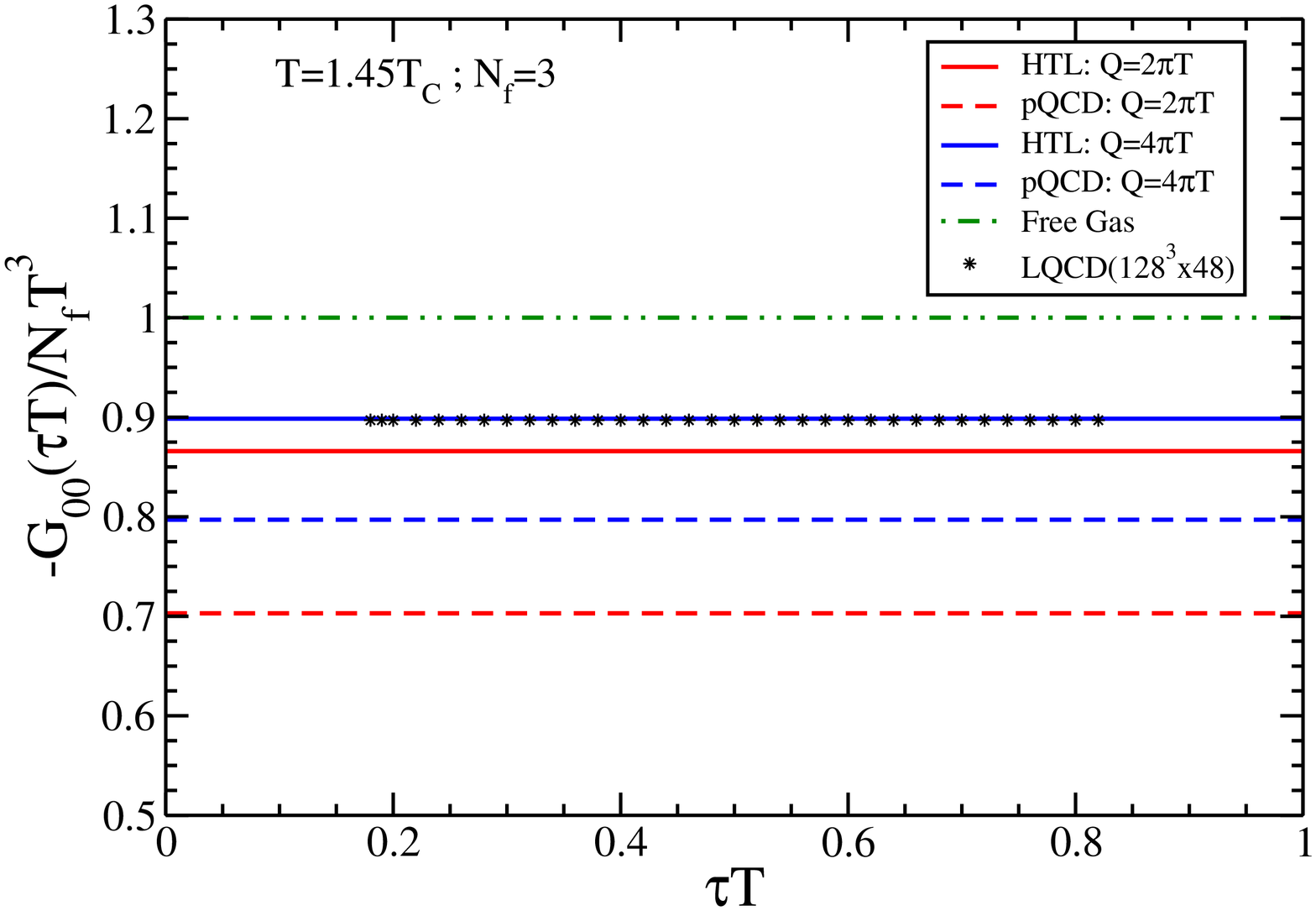}}
%}
%\vspace*{-0.22in}
\caption{(Color online) The scaled  temporal
correlation function with $T^3$ for $N_f=1$ (left panel) and
$N_f=3$ (right panel) at $T=1.45T_C$ for $Q=2\pi T$ (red) and 
$4\pi T$ (blue) as a function of scaled Euclidian time, $\tau T$. 
The symbols represent the recent lattice data~\cite{karsch2} on 
lattices of size $128^3\times 48$ for quark mass $0.1T$ in quenched
QCD.}
\label{corr1}
\end{figure}
 
\begin{figure}[!tbh]
%\subfigure{
%{\includegraphics[scale=0.42]{M_chi.ps}}
{\includegraphics[height=0.48\textwidth, width=0.55\textwidth,
angle=270]{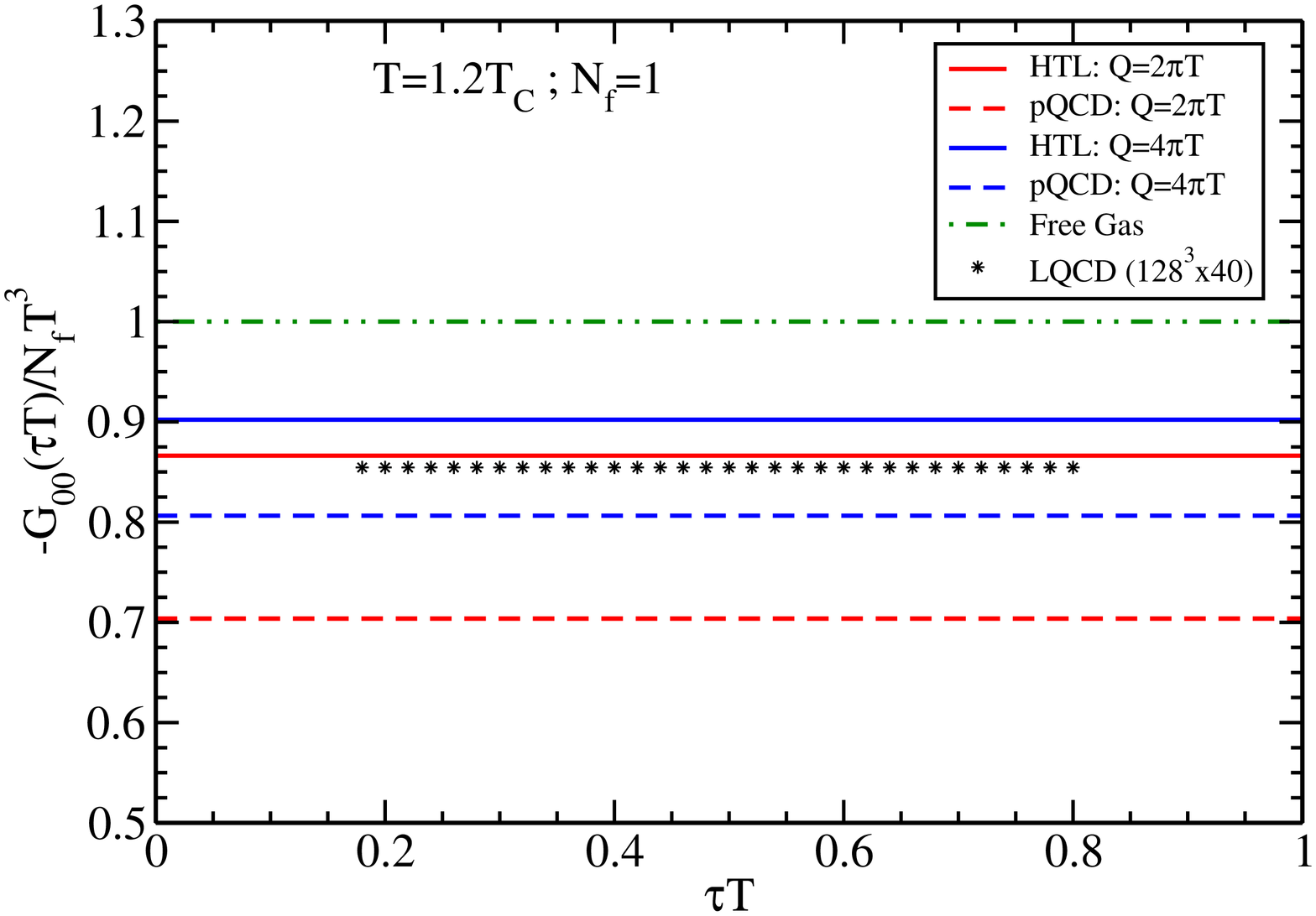}}
%}
%\vspace*{0.3in}
%\subfigure{
%{\includegraphics[scale=0.42]{M_R.ps}}
{\includegraphics[height=0.48\textwidth, width=0.55\textwidth,
angle=270]{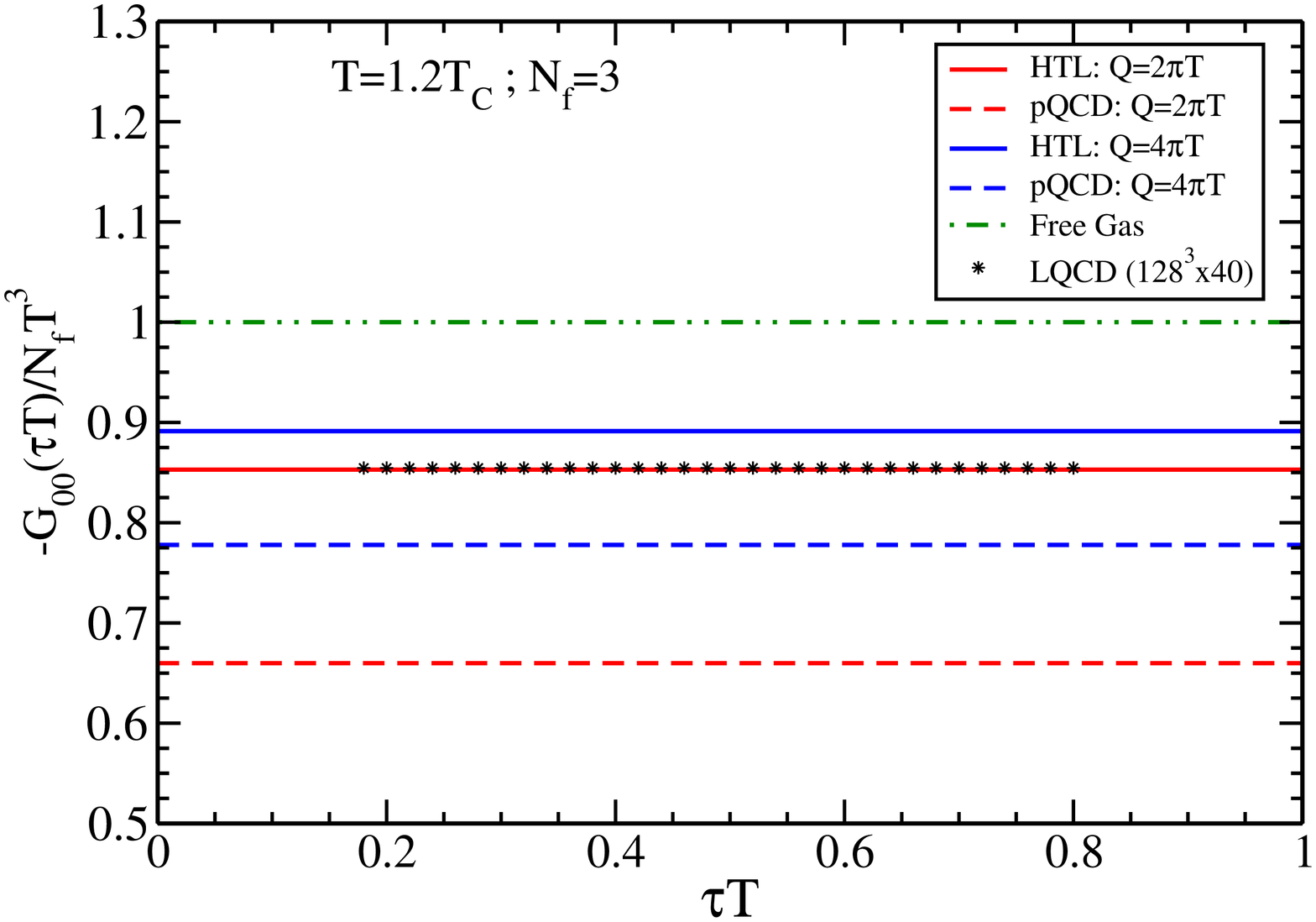}}
%}
%\vspace*{-0.22in}
\caption{(Color online) Same as Fig.~\ref{corr1} but at
$T=1.2T_C$ and the corresponding lattice data are preliminary~\cite{karsch3}
 with
lattice size $148^3\times 40$.}
\label{corr2}
\end{figure}

\begin{figure}[!tbh]
%\subfigure{
%{\includegraphics[scale=0.42]{M_chi.ps}}
{\includegraphics[height=0.48\textwidth, width=0.55\textwidth,
angle=270]{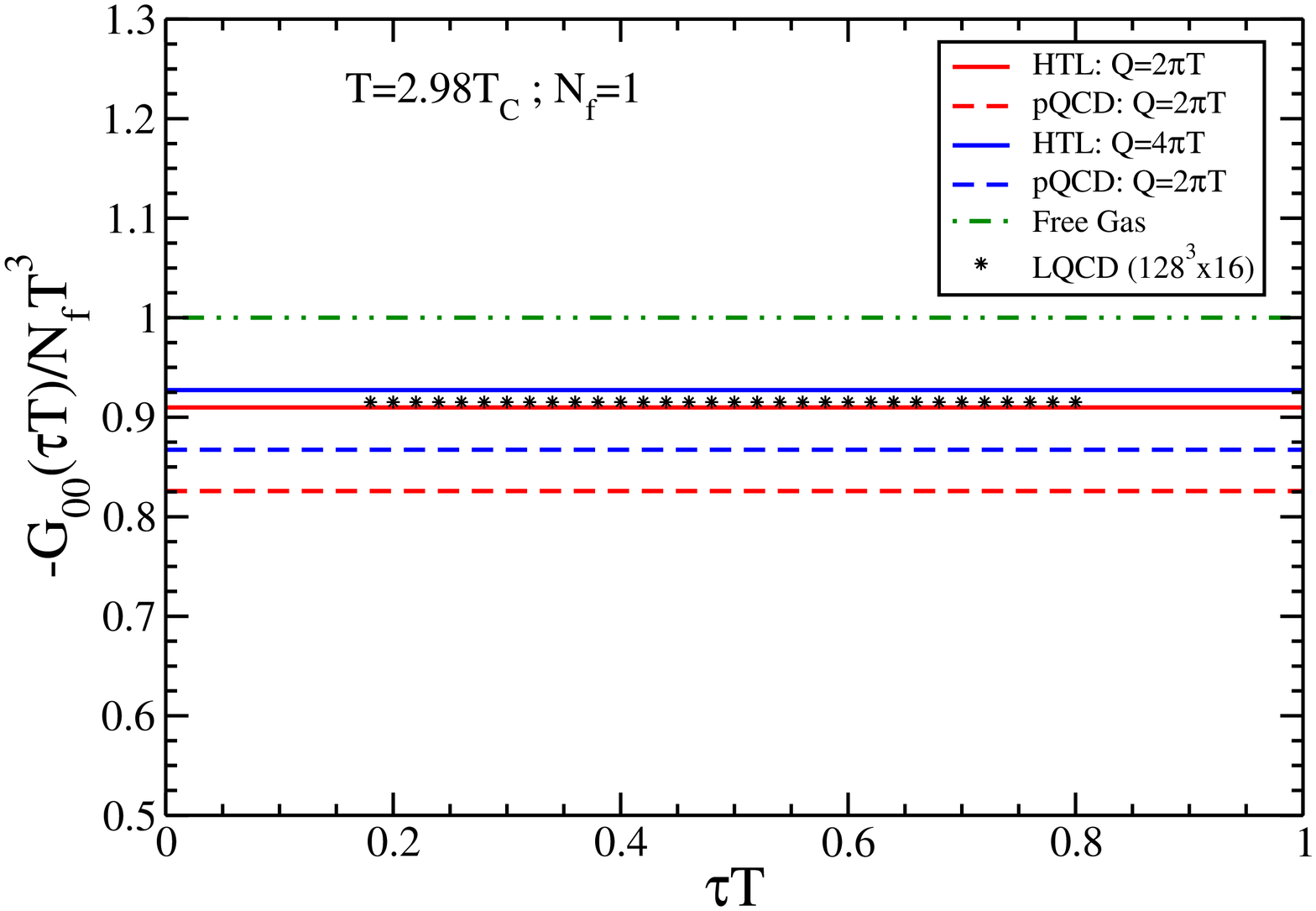}}
%}
%\vspace*{0.3in}
%\subfigure{
%{\includegraphics[scale=0.42]{M_R.ps}}
{\includegraphics[height=0.48\textwidth, width=0.55\textwidth,
angle=270]{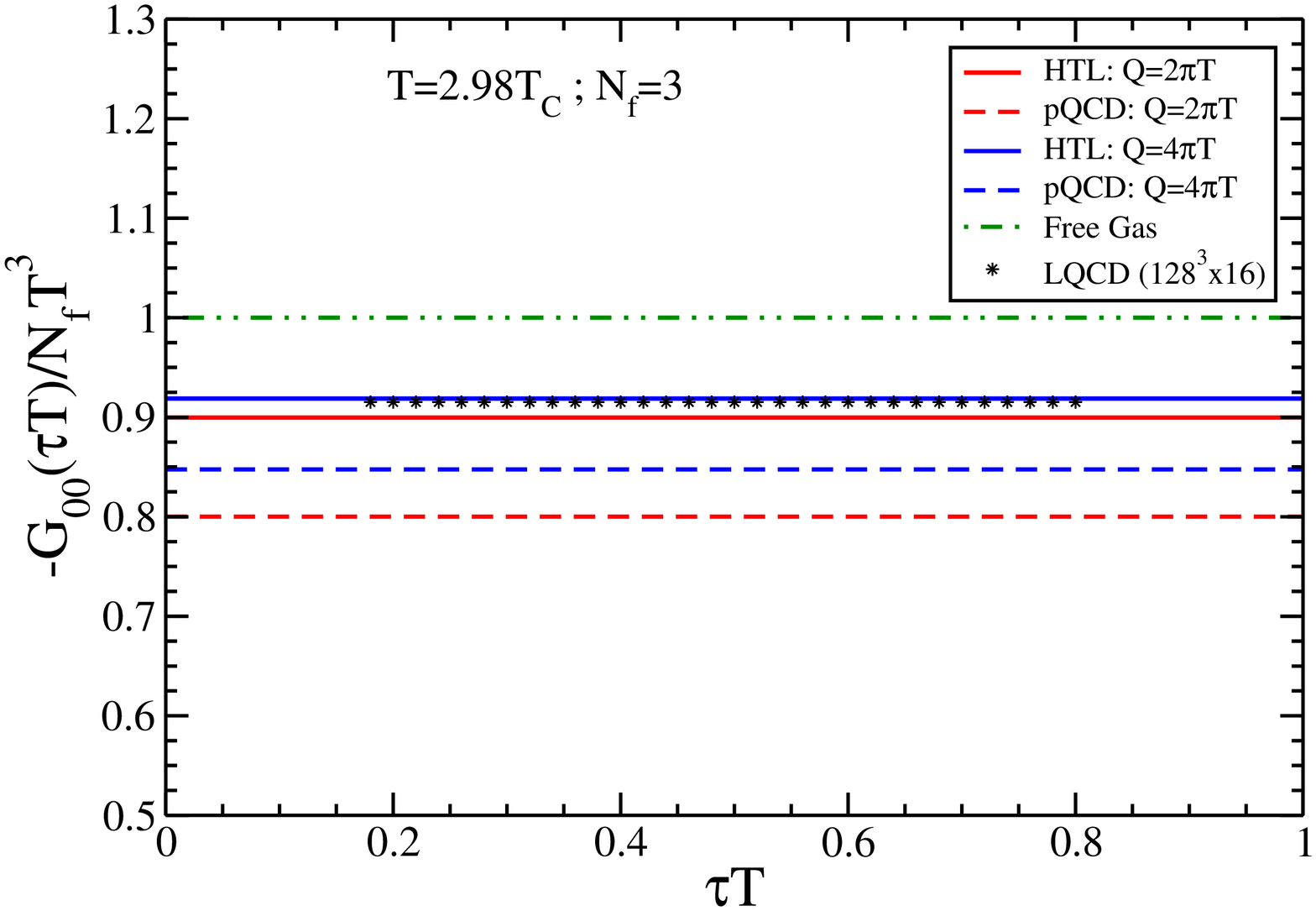}}
%}
%\vspace*{-0.22in}
\caption{(Color online) Same as Fig.~\ref{corr1} but at
$T=2.98T_C$ and the corresponding lattice data are preliminary~\cite{karsch3}
 with lattice size $148^3\times 16$.}
\label{corr3}
\end{figure}

Recently, an improved lattice calculation~\cite{karsch2} has been 
performed within the quenched approximation of QCD where the temporal 
correlation 
function is determined to better than $1\%$ accuracy. 
Using the LO HTLpt QNS in (\ref{s5}) we now obtain the temporal
correlation function in (\ref{eq7}) and compare with the recent lattice 
data~\cite{karsch2}. In Fig.~\ref{corr1} the scaled temporal correlation
function with $T^3$ is shown for $N_f=1$ (left panel) and $N_f=3$ (right panel)
at $T=1.45T_c$. We first note that the correlation functions both in 
HTLpt and pQCD have weak flavour dependence due to the temperature dependent 
coupling, $\alpha_s$ as discussed before. The LO HTLpt result indicates 
an improvement over that of the pQCD one~\cite{blaizot,toimela,kapusta} for 
different choices 
of the renormalisation scale as shown in Fig.~\ref{corr1}. Also, the HTLpt 
result shows a good agreement
to that of recent lattice gauge theory calculation~\cite{karsch2} performed 
on lattices up to size $128^3\times 48$ in quenched approximation for a quark 
mass $\sim 0.1T$. We also note that unlike the 
dynamical spatial part of the correlation function in the vector channel 
the temporal part does not encounter any infrared problem in the 
low energy part 
as it is related to the static quantity through the thermodynamic sum rule
associated with the corresponding symmetry, {\it viz.}, the number conservation of
the system.  We also presented two extreme cases of HTLpt temporal
correlation function at $T=1.2T_C$ in Fig.~\ref{corr2} and at $T\sim 3T_C$ 
in Fig.~\ref{corr3}, respectively, for two different flavours and
compare with the corresponding preliminary lattice data~\cite{karsch3}, which 
are also found to be in good agreement. Finally, we also note that 
even if one compares improved lattice action (asqtad) data~\cite{milc}  
and recent quenched data~\cite{karsch2} for QNS, the quantitative difference 
is within $5\%$ in the temperature domain $T_C\leq T\leq 3T_C$.

%\vspace{0.2in}

%\vspace{0.2in}

\section{Conclusion} 

The LO QNS as a response of the conserved density fluctuation in HTLpt when 
compared with the available lattice data with improved lattice 
actions~\cite{lat3,peter,bazavov,milc,lat2}
in the literature within their wide variation shows the same trend 
but deviates from those in certain extent. The same HTL QNS is used to
compute the temporal part of the Euclidian correlation in the vector current
which agrees quite well with that of improved lattice gauge theory 
calculations~\cite{karsch2,karsch3} recently performed within the quenched 
approximation on lattices up to size $128^3\times48$ for a quark mass 
$\sim 0.1T$. It is also interesting to note that the quantitative difference
between the recent quenched approximation data~\cite{karsch2,karsch3} and
the full QCD data with improved (asqtad) lattice action~\cite{milc} for QNS
is within $5\%$ in the temperature range $T_c\le T\le 3T_c$.  Leaving aside 
the difference
in ingredients in various lattice calculations, one can expect that the HTLpt 
and lattice calculations are in close proximity for quantities associated 
with the conserved density fluctuation.

\begin{acknowledgments}
We are thankful to P. Petreczky for providing us the lattice data on QNS in
Fig.~\ref{qns_I} and
also for clarifying the details of those data. We are also thankful to
F. Karsch for supplying us the preliminary lattice data in Figs.~\ref{corr2}
and \ref{corr3}, and for very useful discussions. MGM is thankful to 
A. De for useful discussions.
\end{acknowledgments}

\end{document}